\documentclass[reprint, superscriptaddress, amsmath,amssymb, aps, prd, nofootinbib]{revtex4-2}

\usepackage{graphicx}
\usepackage{amsmath}
\usepackage{gensymb}
\usepackage{epigraph}
\usepackage{booktabs}
\usepackage{placeins}
\usepackage{soul}
\usepackage[normalem]{ulem}
\usepackage[math]{cellspace}

\usepackage{xspace}
\usepackage[dvipsnames]{xcolor}
\usepackage{fontawesome}
\usepackage{bm}
\usepackage{comment}

\usepackage{hyperref}
\hypersetup{colorlinks=true, allcolors=BrickRed}

\newcommand{\scenario}[1]{\ifcase#1\or Current\or Intermediate\or Optimistic\fi}

\begin{document}

\title{Quasars as a self-sufficient probe of the distance duality relation}

\author{Davide Pedrotti}
\email{davide.pedrotti-1@unitn.it}
\affiliation{Department of Physics, University of Trento, Via Sommarive 14, 38123 Povo (TN), Italy}
\affiliation{Trento Institute for Fundamental Physics and Applications (TIFPA)-INFN, Via Sommarive 14, 38123 Povo (TN), Italy}

\author{J\'ulia Laguna-Miralles}
\affiliation{Institute of Astronomy, University of Cambridge, Madingley Road, CB3 0HA, United Kingdom}
\affiliation{Kavli Institute for Cosmology, University of Cambridge, Madingley Road, CB3 0HA, United Kingdom}

\author{Natalie B. Hogg}
\affiliation{Institute of Astronomy, University of Cambridge, Madingley Road, CB3 0HA, United Kingdom}
\affiliation{Kavli Institute for Cosmology, University of Cambridge, Madingley Road, CB3 0HA, United Kingdom}
\affiliation{Clare Hall, Herschel Road, Cambridge, CB3 9AL, United Kingdom}

\date{\today}

\begin{abstract}
\noindent The distance duality relation links luminosity and angular diameter distances under standard photon propagation. It is usually tested by combining distances from different probes, which requires matching their redshifts and assuming sky isotropy. Self-sufficient probes, which measure both distances for the same object, avoid the first requirement and make the second testable. We assess whether quasars can serve as such probes, combining luminosity distances from the UV--X-ray relation with angular diameter distances from spectroastrometry and reverberation mapping (SARM). As a proof-of-concept, we apply the test to the quasar Mrk~509 and find $\eta\equiv d_{\rm L}/[(1+z)^2d_{\rm A}] = 0.50^{+0.61}_{-0.27}$, consistent with no violation at $0.8\sigma$. We then forecast the number of quasars needed to detect a constant and a linearly redshift-dependent violation, motivated by the findings of Teixeira \textit{et al.} (2025) with potential implications for the Hubble tension and dark energy. A constant violation, $\alpha_0=-0.075$, equivalent to a recalibration of the SNIa distance scale, could be detected at $3\sigma$ with 274--840 quasars. Since SARM distances, unlike BAO, do not rely on early-time physics, this would separate an early-time origin of the Hubble tension from a SNIa miscalibration or a distance duality violation. The redshift dependence is harder to detect: 1000 quasars would exclude $\eta=1$ at $3.4$--$6.6\sigma$, but detecting the dependence itself requires $2.6\times10^4$--$8\times10^4$. GRAVITY+ could bring quasar tests of a constant violation within reach.
\end{abstract}

\maketitle
\section{Introduction}
\label{sec:introduction}
The Etherington reciprocity theorem \citep{Etherington}, also known as the distance duality relation, is the reason why luminosity distances scale as $(1+z)^2$ with angular diameter distance,
\begin{equation}
    d_{\rm L}(z) = (1+z)^2\; d_{\rm A}(z). \label{eq:ddr}
\end{equation}
The validity of this relationship is based on three assumptions: that photon number is conserved, that photons travel on null geodesics, and that spacetime is described by a pseudo-Riemannian metric. 

Under standard metric photon propagation, its most natural physical interpretation is a test of cosmic transparency, where a measurable violation of this relation would signal the presence of exotic physics such as photon--axion mixing \citep{Csaki:2001yk, Bassett:2003vu, Avgoustidis:2010ju}, or a non-minimal coupling between electromagnetism and a scalar field, which would result in a varying of the fine-structure constant \citep{Hees:2014lfa}; or more prosaic phenomena such as scattering by dust in the intergalactic medium. Beyond standard metric photon propagation, violations can also arise if spacetime is not purely Riemannian: for example, in Einstein--Cartan theory, spacetime torsion induces a small but cumulative deviation from the distance duality relation along the photon path \citep{Yun:2026mig}.

Any detected violation of the distance duality relation would have important implications for inference of the Hubble parameter at redshift zero, $H_0$ \citep{Teixeira:2025czm}, and by extension, the Hubble tension. The Hubble tension is inherently a disagreement in distance measurements based on, on the one hand, Cepheid-calibrated luminosity distances from Type Ia supernovae (SNIa) \citep{Riess:2021jrx}; and on the other, angular diameter distances inferred from the cosmic microwave background (CMB) \citep{Planck:2018nkj} and baryon acoustic oscillations (BAO) \citep{DESI:2025zgx} -- see \citet{DiValentino:2021izs} for a review. 

Similarly, the preference for dynamical dark energy reported by the Dark Energy Spectroscopic Instrument (DESI) emerges only when the DESI BAO measurements are combined with CMB and SNIa data \citep{DESI:2025zgx}, and so implicitly assumes that angular and luminosity distances are related by the distance duality relation. For example, \citet{Afroz:2025iwo} found that the Pantheon+ SNIa and DESI DR2 BAO distances are inconsistent with the distance duality relation, with a redshift-dependent signature that mimics dynamical dark energy when the two datasets are combined; accounting for this inconsistency removes the preference for dynamical dark energy. However, when the luminosity and angular diameter distances come from separate probes, it is not possible to determine whether such a violation reflects new physics or unmodelled systematics in one of the datasets. 

To detect violations of the distance duality relation, measurements of luminosity and angular diameter distances are needed, with the canonical examples in cosmology being SNIa \citep{Branch1993} and BAO \citep{SDSS:2005xqv}. Other works have proposed using gravitational waves from standard sirens \citep{Hogg:2020ktc, DeLeo:2025rhy}; HII galaxies \citep{Liu:2021fka, Zheng:2025cgq}; gamma-ray bursts \citep{Xie:2025yun, Alfano:2026aiv, Wu:2026klh}; megamasers \citep{Kanodia:2025jqh}; and cosmic chronometers \citep{Barua:2025dxe}. However, SNIa and BAO measurements are unlikely to cover exactly the same range of redshifts. For example, the angular diameter distances derived from the DESI DR2 BAO lie at seven effective redshifts between $z=0.295$ and $z=2.33$ \citep{DESI:2025zgx}, whereas the Pantheon+ compilation contains 1550 SNIa spread over $0.001 < z < 2.26$ \citep{Scolnic_2022, Brout_2022}. This means that any constraint on the violation of the distance duality relation from these data must be made by interpolating the luminosity distances to the BAO redshifts and extrapolating out to the highest-redshift BAO measurement, which lies beyond the entire supernova sample. This is done either by making assumptions about the underlying cosmology or by employing machine learning \citep{Keil:2025ysb} or Gaussian process reconstruction \citep{Li_2018} methods.

To avoid this, a `matched pair' technique can be adopted, selecting measurements of luminosity and angular diameter distances from different probes at the same redshift; this was done with SNIa and galaxy clusters by \citet{Hu:2026yda}. However, tests of the distance duality relation based on measurements of luminosity and angular diameter distances for different probes inherently rely on assumptions of isotropy. Violations of the distance duality relation caused, for instance, by photo-extinction by intergalactic dust, could depend on the sky direction, since different line-of-sight directions may have different environments \citep{Lin:2020vqj}. 

This problem could be addressed by `self-sufficient' probes of the distance duality relation, namely single objects for which both luminosity and angular diameter distances can be measured simultaneously. When the two distances are obtained from different probes, such as SNIa and BAO, isotropy must be assumed in order to compare them; BAO distances in particular, are essentially averages over large areas of the sky. A self-sufficient probe instead tests the distance duality relation along each individual line of sight to the given object. Combining many such objects, as we propose to do in this work, still assumes that the distance duality relation violation is isotropic, but this assumption becomes testable, for example by splitting the sample by sky region and searching for a dipole in the violation. 

Some proposed self-sufficient probes are strong gravitationally lensed SNIa \citep{Renzi:2020bvl} and strongly lensed gravitational waves \citep{Arjona:2020axn}. However, these events are rare, with strongly lensed gravitational waves having not yet even been observed, which motivates us to consider alternative avenues. In this paper, we propose the use of quasars.

A quasar is a highly luminous active galactic nucleus (AGN), driven by the accretion of gas onto the central supermassive black hole \citep{Padovani:2017zpf}. Their brightness means they are observable at extremely high redshifts, e.g. $z=7.5$ \citep{Lusso_2020} to $z=10$ \citep{Whalen2023}, and their luminosities can be standardised \citep{Lusso_2017, Lusso_2020, Sacchi:2022ofz}. Furthermore, angular diameter distances to them can be measured, either from the empirically calibrated sizes of compact radio cores \citep{Kellermann_1993, Gurvits_1999, Cao_2017}, or geometrically, by combining spectroastrometry with reverberation mapping \citep{Elvis:2002ja, Wang:2019gaq, Li_2022, Li_2025}, making quasars a powerful probe of cosmology.

\citet{Zheng:2020fth} investigated the use of quasars to constrain the distance duality relation, utilising separate quasar populations to measure luminosity and angular distances. In this work, in contrast, we investigate the possibility of measuring both luminosity and angular diameter distances from the same individual quasars, via the UV--X-ray luminosity relation, and spectroastrometry combined with reverberation mapping (SARM), respectively. Although SARM measurements are currently restricted to low redshifts $z \lesssim 0.2$ \citep{Li_2025}, future experiments are expected to extend such measurements to $z \sim 2$--$3$ \citep{Amorim_2021, GRAVITYplusWhitePaper}. As a proof-of-concept, we first apply this test to Mrk~509~\citep{Markaryan1973}, currently the only quasar with both a SARM angular diameter distance and a UV--X-ray luminosity distance. Notably, while single-object tests of the distance duality relation have been conducted previously, for instance, using the SZ/X-ray gas fractions of galaxy clusters \citep{Holanda:2012at}, our test on Mrk~509 is, to our knowledge, the first based on independent measurements of both the luminosity and angular diameter distances for the same source. We then forecast the constraints on distance duality relation violations achievable with future quasar datasets. We build mock catalogues of luminosity and angular diameter distances, and use these to infer constraints on parametric forms of the distance duality violation motivated by the Hubble tension and the evidence for dynamical dark energy in recent cosmological observations \citep{Teixeira:2025czm}. The use of the same quasars to measure both luminosity and angular diameter distances means that the constraints on the distance duality relation will be independent of any cosmological parameters; they will, however, depend on the relative calibration of the two distance scales, as we discuss further below.

The paper is structured as follows: in \autoref{sec:quasars}, we describe how distances can be measured from quasars; in \autoref{sec:method} we describe our forecasting method; in \autoref{sec:results} we present the Mrk~509 measurement and our forecast results; and we conclude with \autoref{sec:conclusions}.

\section{Distances from quasars}
\label{sec:quasars}

\subsection{Luminosity distances}
\label{sec:quasarsasstandardcandles}

Unobscured (Type~I) quasars, i.e. luminous AGN viewed such that the broad-line region (BLR) is directly visible, can be used as standardisable candles: their luminosities can be inferred statistically from empirical relations between intrinsic luminosity and observables connected to the accretion flow. The most widely used method for obtaining quasar luminosity distances is based on the non-linear relation between the rest-frame monochromatic luminosities at $2\,{\rm keV}$ and $2500\,\text{\AA}$ \citep{Risaliti_2015, Risaliti_2019},
\begin{equation}
  \log L_{\rm X} = \gamma \log L_{\rm UV} + \beta ,
  \label{eq:uvx_luminosity_relation}
\end{equation}
where $\gamma$ is the slope of the UV--X-ray relation, $\beta$ the intercept, and $\log$ denotes $\log_{10}$ throughout. Writing each luminosity as $L = 4\pi d_{\rm L}^2 F$, in terms of the rest-frame (K-corrected) fluxes $F_{\rm X}$ and $F_{\rm UV}$, and solving for the luminosity distance gives

\begin{equation}
  \log d_{\rm L}
  =
  \frac{\gamma \log F_{\rm UV} - \log F_{\rm X} + \beta_F}{2(1-\gamma)} ,
  \label{eq:uvx_distance}
\end{equation}
where $\beta_F$ absorbs $\beta$ and the normalisation constants. This relation allows each quasar with suitable optical/UV and X-ray data to provide an estimate of \(d_{\rm L}\), once the intercept has been calibrated.

The full UV--X-ray quasar catalogue of \citet{Lusso_2020} contains 2421 sources over \(0.009\leq z\leq7.541\). For cosmological applications, however, the clean low-redshift end is more restricted: \citet{Lusso_2025} use 2036 quasar measurements, consisting
of 2023 quasars at \(z>0.7\) plus 13 local AGN at \(0.009<z<0.087\) with direct UV spectroscopy. The intermediate low-redshift range is conservatively excluded unless rest-frame
\(2500\,\text{\AA}\) can be measured directly, because host-galaxy light can contaminate the optical continuum and bias the inferred UV luminosity. For comparison with angular-diameter-distance probes, the robust UV--X-ray luminosity-distance sample should thus be regarded as mainly covering \(z\gtrsim0.7\), with a small local calibration anchor, and extending to \(z\simeq7.5\) \citep{Risaliti_2015,Risaliti_2019,Lusso_2020,Lusso_2025}. 

In the overlap interval used for cross-calibration with Type~Ia supernovae, \(0.7<z<1.6\), the inferred quasar distances are broadly consistent with the supernova Hubble diagram; at \(z\gtrsim1.5\)--\(2\), several analyses have reported departures from flat \(\Lambda\)CDM within the adopted parametrisations, motivating continued scrutiny of both cosmology and possible residual quasar systematics \citep{Risaliti_2019, Lusso_2020,Lusso_2025}.

The error budget on the luminosity distance inferred by this method is dominated by the scatter in the UV--X-ray relation. Propagating the scatter through Eq.~\eqref{eq:uvx_distance}, the luminosity distance uncertainty for quasar $i$ is
\begin{equation}
  \sigma_{\log d_{\rm L},i}^{2}
  =
  \frac{\sigma_{\rm UVX}^{2} + \sigma_{\log F_{\rm X},i}^{2} + \gamma^{2}\sigma_{\log F_{\rm UV},i}^{2}}{[2(1-\gamma)]^{2}},
  \label{eq:uvx_distance_uncertainty}
\end{equation}
where $\sigma_{\rm UVX}$ is the intrinsic scatter of the relation. We assume independent flux errors and keep $\gamma$ and $\beta$ fixed. \citet{Lusso_2020} find $\gamma = 0.600 \pm 0.015$, with no significant evolution in redshift. The factor $2(1-\gamma) \approx 0.8$ amplifies the scatter: the total residual of $0.24$ dex around the UV--X-ray relation in current samples (E. Lusso, private communication) becomes $0.3$ dex in distance, or $1.5$ mag in distance modulus, compared with $\simeq0.15$ mag for standardised SNIa \citep{Brout_2022, Scolnic_2022}.

Other luminosity distance indicators for quasars have also been proposed, such as the C~{\sc iv} Baldwin effect \citep{Huang_2023CIV} and AGN variability. The latter already achieves a smaller scatter of $0.81$ mag on the Hubble diagram \citep{Dutra_2025} and could be applied to $10^5$--$10^6$ AGN in upcoming time-domain surveys \citep{LagunaMiralles_2026}. In this work, we restrict ourselves to the UV--X-ray relation, which is currently the most mature probe.

\subsection{Angular diameter distances}
Angular diameter distances to quasars have in the past been obtained by treating the compact radio cores, observed with Very Long Baseline Interferometry (VLBI), as standard rulers \citep{Kellermann_1993, Gurvits_1999, Cao_2017}, and these angular distances have been used to test the distance duality relation \citep{Li_2018}. However, VLBI samples generally consist of radio-loud, jet-dominated quasars, whilst the luminosity distance method described in \autoref{sec:quasarsasstandardcandles} above relies on radio-quiet, unobscured Type~I quasars \citep{Huang_2024RLQ}. The two methods therefore cannot be applied to the same individual objects, which is the goal of this paper. We thus consider a geometric method which in principle can be applied to the same quasars for which luminosity distances can be obtained: spectroastrometry combined with reverberation mapping (SARM).

Reverberation mapping measures the physical size of the BLR from the time lag between variations in the continuum and in a broad emission line \citep{Peterson_2004, Bentz_2013},
\begin{equation}
    R_{\rm BLR} = c\tau_{\rm rest} = \frac{c\tau_{\rm obs}}{1+z},
  \label{eq:rblr_rm}
\end{equation}
while spectroastrometry measures its angular size, $\theta_{\rm BLR}$, from the wavelength-dependent centroid shift across the broad line. Their ratio gives a geometric angular diameter distance~\citep{Elvis:2002ja, Wang:2019gaq},
\begin{equation}
  d_{\rm A}(z) = \frac{R_{\rm BLR}}{\theta_{\rm BLR}} ,
  \label{eq:sarm_da}
\end{equation}
which requires no calibration against other distance indicators, although it does depend on modelling of the BLR geometry and kinematics.

The feasibility of this method has been demonstrated observationally. Following the detection of BLR rotation in 3C~273 using the GRAVITY instrument of the Very Large Telescope Interferometer (VLTI) \citep{GRAVITY_2018}, a SARM analysis of this object yielded the first geometric distance to a quasar, $d_{\rm A}=551.5^{+97.3}_{-78.7}\,{\rm Mpc}$ \citep{Wang:2019gaq} (subsequently revised to $d_{\rm A}=676^{+737}_{-321}\,{\rm Mpc}$ by \citet{Li_2022}); see also \citet{Amorim_2021} for NGC~3783. \citet{Li_2025} have since applied the method to four quasars at $z<0.19$, with uncertainties of $0.23$--$0.52$ in $\ln d_{\rm A}$, obtaining $H_0=69^{+12}_{-10}\,{\rm km\,s^{-1}\,Mpc^{-1}}$. 

Current SARM measurements are therefore proof-of-concept rather than precision cosmology, since the sample is still small and the observations require high signal-to-noise
reverberation mapping together with microarcsecond-scale angular information. As we noted in \autoref{sec:introduction}, Mrk~509 is common to the SARM sample of \citet{Li_2025} and the UV--X-ray sample of \citet{Lusso_2020}, providing the measurements needed for our proof-of-concept test in \autoref{sec:results}.  

Upgrades to GRAVITY are expected to extend SARM to several hundred Type~I AGN up to $z\sim2$--$3$, with $\mathcal{O}(10\%)$ uncertainties on $d_{\rm A}$ \citep{Amorim_2021, GRAVITYplusWhitePaper}; simulations suggest that $\sim60$ such objects would constrain $H_0$ to $\sim2\%$ \citep{Songsheng_2021}. BLRs have already been spatially resolved at $z\simeq2.3$ and $z\simeq4$ \citep{Abuter_2024, GRAVITYplus2026J0529}, although distances have not yet been measured for these objects. For now, however, the angular diameter distance is the bottleneck for a quasar-only test of the distance duality relation: the current SARM catalogues contain around 500 times fewer objects than UV--X-ray catalogues, and are confined to $z<0.2$. In the following sections, we therefore explore the required size and precision of a future quasar sample that would be needed to obtain constraints on distance duality relation violations of the same magnitude as that reported by \citet{Teixeira:2025czm}.

\section{Method}
\label{sec:method}

We generate mock catalogues of quasar distance observations assuming a fiducial cosmology and distance duality relation model, and use them to forecast constraints on the violation parameters. We use \autoref{eq:ddr} to define
\begin{equation}
    \eta(z) = \frac{d_{\rm L}(z)}{(1+z)^2 d_{\rm A}(z)},  \label{eq:eta}
\end{equation}
such that deviations from $\eta(z)=1$ correspond to a violation of the distance duality relation. Following \citet{Teixeira:2025czm}, we then parameterise $\eta(z)$ using the following models: a constant offset, $\eta(z) = 1+\alpha_0$ (M1), and a linear redshift evolution, $\eta(z) = 1+\alpha_0+\alpha_1 z$ (M2). In each case, $\alpha_0, \alpha_1 \neq 0$ indicates a violation of the distance duality relation. 

We construct mock catalogues of $N$ quasars with redshifts evenly spaced over $0.2\leq z\leq2.0$. This choice is motivated both by the comparison we will make with the results of \citet{Teixeira:2025czm}, meaning that our mocks must cover approximately the same range as the DESI BAO angular diameter distance measurements in order to be comparable; and by the fact that future SARM measurements by GRAVITY+ are expected to reach $z \sim 2$--$3$ \citep{Amorim_2021, GRAVITYplusWhitePaper}, making our range a reasonably conservative choice. The mock angular diameter distances are computed using the fiducial cosmology, and the corresponding luminosity distances are then obtained via the selected distance duality relation model,
\begin{equation}
  d_{\rm L}^{\rm true}(z_i) = \eta(z_i)(1+z_i)^2 d_{\rm A}^{\rm true}(z_i).
\end{equation}

We model the observational errors as Gaussian in the logarithm of each distance, with standard deviations $\sigma_{\ln d_{\rm L}}$ and $\sigma_{\ln d_{\rm A}}$, independent between the two distances and between quasars. Rather than drawing random realisations, we use noise-free catalogues in which each mock distance equals its fiducial value; the resulting posterior is the expected posterior for $N$ quasars, free of the scatter of any single realisation. 

We consider three scenarios for the per-object scatter, summarised in \autoref{tab:scenarios}. For the luminosity distances, the individual flux uncertainties of \citet{Lusso_2020} are small compared with the dispersion of the UV--X-ray relation, meaning that we neglect the redshift dependence and assume the scatter to be constant, $\sigma_{\ln d_{\rm L}}=\ln(10)\,\sigma_{\rm UVX,eff}/[2(1-\gamma)]$, where $\sigma_{\rm UVX,eff}$ is the total dispersion of the relation, including flux errors. 

Our ``\scenario1'' value of $\sigma_{\rm UVX,eff}= 0.24$ dex is the total residual of current samples given in \autoref{sec:quasarsasstandardcandles}, while $0.17$ dex is potentially achievable, and $0.14$ dex is a practical floor below which the remaining sample would be unsuitable for cosmology (E.~Lusso, private communication; see also \citep{Signorini:2023nyd}). For the angular diameter distances, our \scenario1\footnote{In this case the word \textit{Current} is a slight abuse of notation, as here we are extrapolating the \textit{current} uncertainties on $d_{\rm A}$ measurements to much higher redshifts.} value of $\sigma_{\ln d_{\rm A}}=0.30$ is representative of the four SARM measurements of \citet{Li_2025}, while our ``\scenario2'' and ``\scenario3'' values of $0.20$ and $0.15$ are chosen to be reasonably conservative improvements on the Current value relative to the $\sim10\%$ precision anticipated with GRAVITY+ \citep{Amorim_2021}. Here we are implicitly assuming that uncertainties on $d_{\rm A}$ can be modelled as constant functions of redshift in the three scenarios studied. We defer redshift modelling of $d_{\rm A}$ uncertainties to future studies, when more SARM data will be available.

\begin{table}
  \begin{tabular}{Slcccc}
      \hline
      Scenario & $\sigma_{\rm UVX,eff}$ [dex] & $\sigma_{\ln d_{\rm L}}$ & $\sigma_{\ln d_{\rm A}}$ & $\sigma_{\ln\eta}$ \\
      \hline
      \scenario1  & 0.24 & 0.69 & 0.30 & 0.75 \\
      \scenario2   & 0.17 & 0.49 & 0.20 & 0.53 \\
      \scenario3 & 0.14 & 0.40 & 0.15 & 0.43 \\
      \hline
  \end{tabular}
\caption{Per-object scatter in each forecast scenario. The final column gives the resulting scatter in $\ln\eta$ (\autoref{eq:ddr_log_variance}).}
\label{tab:scenarios}
\end{table}

Since both distances are measured from the same quasar, each mock object directly yields a measurement, $\eta_{{\rm obs},i}$, obtained by inserting its mock distances into \autoref{eq:eta}, with no interpolation between redshifts required. Treating spectroscopic redshift errors as negligible, the assumed log-distance model gives
\begin{equation}
  \sigma_{\ln\eta}^{2} = \sigma_{\ln d_{\rm L}}^{2} + \sigma_{\ln d_{\rm A}}^{2}.
  \label{eq:ddr_log_variance}
\end{equation}
We therefore use the likelihood
\begin{equation}
\ln\mathcal{L} = -\frac{1}{2}\sum_{i=1}^{N}\frac{\left[\ln\eta_{{\rm obs},i} - \ln\eta(z_i)\right]^{2}}{\sigma_{\ln\eta}^{2}} + {\rm const.}
  \label{eq:ddr_log_likelihood}
\end{equation}
Since $\sigma_{\ln\eta}$ already includes the intrinsic scatter of both distance methods, we do not fit an additional intrinsic scatter parameter. 

We sample the M2 posterior using the MCMC sampler of \texttt{Cobaya} \citep{Torrado:2020dgo, Lewis:2013hha}, with uniform priors $\alpha_0\in[-0.9,1.5]$ and $\alpha_1\in[-1, 1]$, and require $\eta(z)>0$ across the sample. Because the per-object scatter is constant, the sample size required for a given detection significance can also be computed without sampling: analytically for M1 (\autoref{sec:results}), and from the Fisher matrix of \autoref{eq:ddr_log_likelihood} for M2, which reproduces the MCMC uncertainties at $N=1000$. 

The likelihood in \autoref{eq:ddr_log_likelihood} does not depend on cosmological parameters: the fiducial cosmology enters only through the construction of the mocks, for which we assume a Planck 2018-like fiducial cosmology, with $H_0 = 67.36$ km s$^{-1}$ Mpc$^{-1}$ and $\Omega_{\rm m} = 0.314$ \citep{Planck:2018vyg}. However, the constraints \textit{are} conditional on the relative calibration of the two distance scales, since any common multiplicative offset between them is degenerate with $\alpha_0$. We will discuss this point further in \autoref{subsec:calibration}. 

\section{Results and discussion}
\label{sec:results}

\subsection{Mrk 509}
As noted in \autoref{sec:quasars}, Mrk~509 ($z=0.0344$) \citep{Markaryan1973} is the only one of the four SARM quasars of \citet{Li_2025} that also appears in the UV--X-ray sample of \citet{Lusso_2020}. Before examining the forecast future constraints on the distance duality relation with quasars, we can therefore use this quasar to obtain a constraint on $\eta(z=0.0344)$, as a proof-of-concept of our method. 

\citet{Lusso_2020} provide a distance modulus for each quasar in their sample, with the intercept of the UV--X-ray relation cross-calibrated against Type~Ia supernovae assuming $H_0=70\,{\rm km\,s^{-1}\,Mpc^{-1}}$. For Mrk~509, this gives $d_{\rm L}=136.5$ Mpc. The quoted uncertainty includes the flux and calibration errors but not the intrinsic scatter of the relation; adding the intrinsic scatter of $0.21$ dex from \citet{Lusso_2020} gives $\sigma_{\ln d_{\rm L}}=0.61$. For the angular diameter distance, \citet{Li_2025} find $\log(d_{\rm A}/{\rm Mpc})=2.41^{+0.22}_{-0.23}$, corresponding to $d_{\rm A}=257$ Mpc with $\sigma_{\ln d_{\rm A}}=0.52$, after symmetrising the uncertainty on $d_{\rm A}$.

Combining these, we find $\eta = 0.50^{+0.61}_{-0.27}$, consistent with the distance duality relation at $0.8\sigma$. The central value is driven below $\eta=1$ mainly by the SARM distance, which lies above the flat $\Lambda$CDM expectation, albeit within its uncertainty~\citep{Li_2025}. Rescaling the supernova calibration to $H_0=67.36\,{\rm km\,s^{-1}\,Mpc^{-1}}$, as used in our mocks, shifts the result only slightly, to $\eta = 0.52^{+0.63}_{-0.28}$, but illustrates the dependence on the relative calibration of the two distance scales discussed in \autoref{sec:method}. While this single measurement places no meaningful constraint on the violations considered here, it demonstrates that the test can already be performed on real data. Its uncertainty, an order of magnitude larger than the fiducial violation $\alpha_0=-0.075$, also shows why large samples of quasars will be needed to provide precise self-sufficient constraints on distance duality violation, as we now quantify.

\subsection{Forecast future constraints}
\subsubsection{M1}
Firstly, we examine the constraints that could be obtained on the violation parameter $\alpha_0$ in M1, taking as a fiducial value $\alpha_0=-0.075$, as found by \citet{Teixeira:2025czm} from the combination of Planck 2018 \citep{Planck:2018vyg}, DESI BAO \citep{DESI:2024mwx} and Pantheon+ SNIa \citep{Scolnic_2022, Brout_2022} calibrated with a SH0ES prior \citep{Riess:2021jrx}. Since $\eta$ is constant in M1, each quasar provides an independent measurement of $\ln(1+\alpha_0)$ with scatter $\sigma_{\ln\eta}$, so that the uncertainty on $\ln(1+\alpha_0)$ decreases as $\sigma_{\ln\eta}/\sqrt{N}$. Distinguishing the fiducial violation from zero at $3\sigma$ therefore requires
  \begin{equation}
      N \geq \left[\frac{3\,\sigma_{\ln\eta}}{\ln(1+\alpha_0)}\right]^2 .
      \label{eq:N_required}
  \end{equation}

\autoref{tab:M1_N} gives the resulting sample sizes for each error scenario. In the \scenario1 scenario, $840$ quasars are needed; this falls to $414$ and $274$ in the \scenario2 and \scenario3 scenarios, respectively. By contrast, the four quasars with existing SARM distances give an uncertainty of $\simeq0.38$ on $\alpha_0$, five times larger than the fiducial violation. Since the required sample size scales as $\sigma_{\ln\eta}^2$, it is dominated by the luminosity distance scatter: even a perfect angular diameter distance measurement would reduce the requirement in the \scenario1 scenario only from $840$ to $707$ quasars.

\begin{table}
  \centering
  \begin{tabular}{Slc}
      \hline
      Scenario & $N_{3\sigma}$ \\
      \hline
      \scenario1 & 840 \\
      \scenario2 & 414 \\
      \scenario3 & 274 \\
      \hline
  \end{tabular}
  \caption{Number of quasars required to distinguish $\alpha_0=-0.075$ (M1) from zero at 3$\sigma$ in each error scenario.}
  \label{tab:M1_N}
\end{table}

For comparison, GRAVITY+ is expected to provide spectroastrometry for several hundred Type~I AGN \citep{GRAVITYplusWhitePaper}. Assuming the measurement precision of our \scenario3 scenario, a detectable violation of the distance duality relation as found by \citet{Teixeira:2025czm} is within reach, provided that most of the 274 required sources also have reverberation lags and X-ray data suitable to derive luminosity distances via the UV--X-ray relation. If, however, the error budget remains as it is today, in the \scenario1 scenario, a robust $3\sigma$ detection of distance duality violation may not be possible due to the large number of quasars required (840).

\subsubsection{M2}
Secondly, we investigate the M2 model, taking as fiducial values $\alpha_0=-0.070$ and $\alpha_1=-0.014$, as found by \citet{Teixeira:2025czm} for this model, using the same combination of data as in the case of M1 (CMB, BAO, SNIa). Since the two violation parameters are strongly anti-correlated over our redshift range, we fix the sample size at $N=1000$. This choice of sample size is large enough that the constant violation of M1 would be detected at $3\sigma$ in all three error scenarios (\autoref{tab:M1_N}), but is not far beyond what forthcoming instruments such as GRAVITY+ may provide \citep{GRAVITYplusWhitePaper}, as discussed in \autoref{sec:quasars}.

\begin{figure}
  \includegraphics[width=\columnwidth]{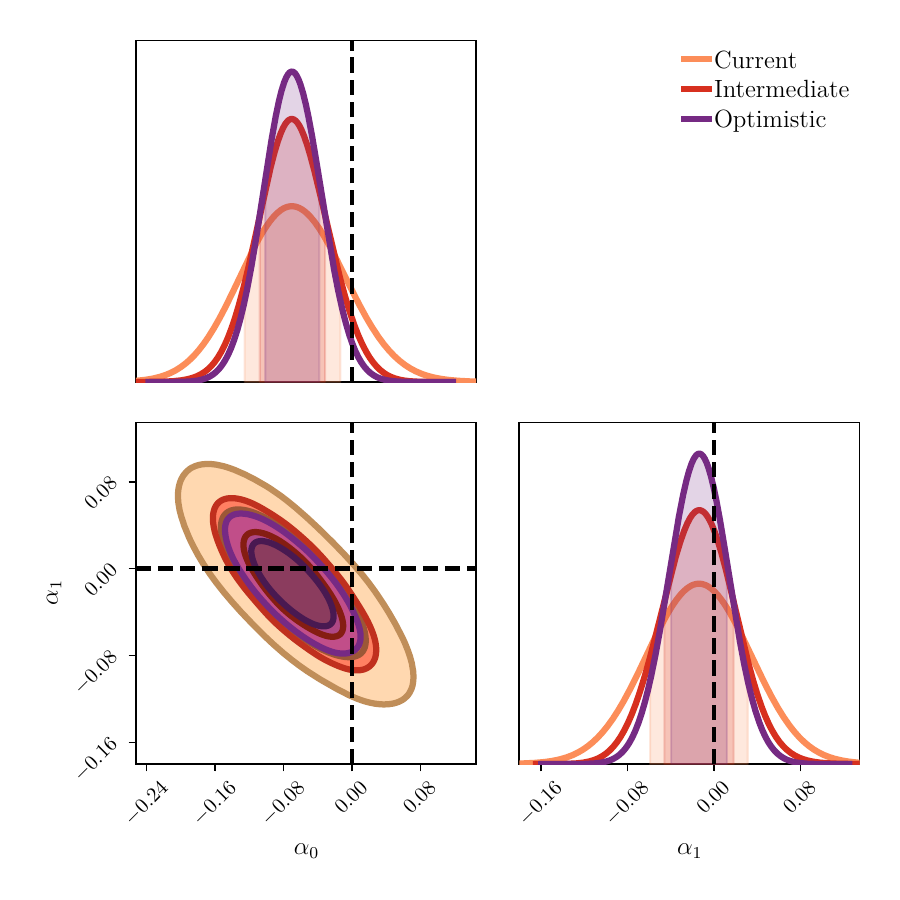}
  \caption{One- and two-dimensional posterior distributions of the M2 parameters $\alpha_0$ and $\alpha_1$ for $N=1000$ quasars in each error scenario. The shaded regions show the 68\% and 95\% confidence intervals, whilst the dashed lines indicate $\alpha_0=\alpha_1=0$. The fiducial values are $\alpha_0=-0.070$ and $\alpha_1=-0.014$.}
  \label{fig:M2}
\end{figure}

The result is shown in \autoref{fig:M2}. Over our redshift range, the data mainly constrain the average violation across the sample, and the same average can be produced by a larger $\alpha_0$ with a smaller $\alpha_1$, or vice versa. As a result, neither parameter is individually inconsistent with zero; in the \scenario1 scenario, for example, we find $\alpha_0=-0.068\pm0.051$ and $\alpha_1=-0.014\pm0.042$. Nevertheless, the case of no violation, $\alpha_0=\alpha_1=0$, lies outside the joint posterior, and is excluded at $3.4\sigma$, $5.2\sigma$ and $6.6\sigma$ in the \scenario1, \scenario2 and \scenario3 scenarios, respectively. In other words, a sample of $1000$ quasars would detect the violation itself, but not its evolution with redshift. If the error budget stays as it is in the \scenario1 scenario, detecting $\alpha_1\neq0$ at $3\sigma$ would require $8\times10^4$ quasars with both UV--X-ray luminosity distances and SARM angular diameter distances; in the \scenario3 scenario, this falls to $2.6\times10^4$. 

\subsection{Calibration dependence of the constraints}
\label{subsec:calibration}

\begin{figure*}
    \centering
    \includegraphics[width=0.75\textwidth]{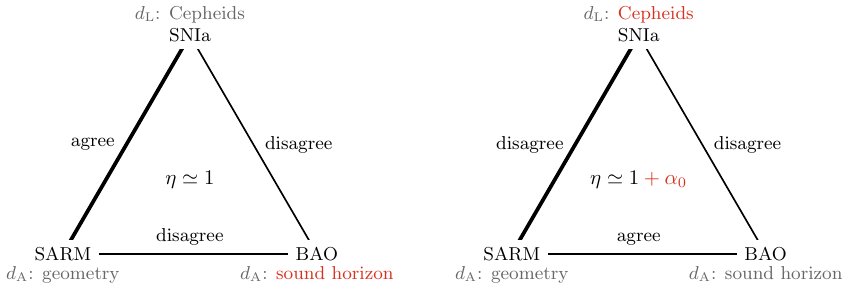}
    \caption{The two possible outcomes of a self-sufficient quasar measurement of $\eta$ (thick line), where quasar luminosity distances have been anchored to SNIa. Each corner of the triangles is a different distance scale; the disagreement between SNIa and BAO is the Hubble tension. 
    \textit{Left}: the quasar luminosity and angular diameter distances agree, so the SNIa scale is consistent with geometry, and the Hubble tension must point to unaccounted-for early-time physics modifying the sound horizon. \textit{Right}: the quasar distances disagree, implying that the Hubble tension arises either from a violation of distance duality, or from a miscalibration of the SNIa. These possibilities are indicated with red text in both panels.}
    \label{fig:triangles}
\end{figure*}

Finally, we discuss the impact of \textit{calibration} of the two distance measurements on the distance duality relation constraints that we have obtained. Interpreting a constant offset in the distance duality relation requires knowledge of the relative calibration of the two distance scales, even though the likelihood in \autoref{eq:ddr_log_likelihood} is independent of cosmology. As mentioned in \autoref{sec:quasarsasstandardcandles}, the UV--X-ray relation yields only relative distances before calibration: \citet{Lusso_2020} fix the intercept $\beta_F$ in \autoref{eq:uvx_distance} by shifting the quasar Hubble diagram onto that of Pantheon SNIa in the common redshift range (Pantheon+ over $0.7<z<1.6$ in \citet{Lusso_2025}), with $H_0$ fixed to $70\,{\rm km\,s^{-1}\,Mpc^{-1}}$. The SARM distances of \citet{Li_2025}, by contrast, require no external anchor. 

A constant relative calibration factor $C$ therefore enters the parametrisation in \autoref{eq:eta} as $\eta_{\rm obs}(z)=C\,\eta(z)$. In M1 it is exactly degenerate with $1+\alpha_0$. This degeneracy is expected given the choice of fiducial model: the constant offset of \citet{Teixeira:2025czm} is, by construction, the rescaling of the SNIa distance scale that restores concordance with BAO calibrated on the CMB sound horizon, i.e.\ the Hubble tension in disguise, as reflected in $\alpha_0\simeq-0.075$ being comparable to the fractional difference between the CMB and SH0ES values of $H_0$, $67.4/73-1\simeq-0.077$. 

Far from being a drawback, this is what makes the constant-offset test interesting for the Hubble tension. The absolute scale of BAO distances relies on the sound horizon, and therefore on early-time physics \citep{SDSS:2005xqv, Aylor:2018drw, Arendse:2019hev}, whereas SARM distances are geometric and independent of it, although they depend on the modelling of the BLR. A quasar measurement of $\eta$, with UV--X-ray distances anchored to the SNIa scale, therefore discriminates between two classes of explanation for the SNIa--BAO mismatch, which we represent schematically in \autoref{fig:triangles}. 

As indicated in the figure, if $\eta\simeq1$, the SNIa scale agrees with the geometric SARM scale and the mismatch must originate on the BAO side, for instance from new physics before recombination modifying the sound horizon (barring a fine-tuned cancellation between a SNIa miscalibration and a violation of the distance duality relation). If $\eta$ is instead consistent with the fiducial offset, the mismatch must originate on the SNIa side, either from a SNIa miscalibration, including late-time effects on the SNIa luminosity scale, or from a genuine violation of the distance duality relation. 

Separating these two effects would require either anchoring the UV--X-ray relation with something other than SNIa, or measuring the redshift dependence of $\eta$. A calibration offset would be constant across all redshifts, whereas violations of distance duality arising from, for instance, photon--axion mixing  \citep{Csaki:2001yk, Bassett:2003vu} or cosmic opacity \citep{Avgoustidis:2010ju}, would be redshift dependent, and should vanish as $z \to 0$. One alternative to SNIa would be to anchor the UV--X-ray luminosity distance relation to angular distances from the low-redshift SARM quasars. This sets $\eta=1$ at low redshift by construction, and in turn means that the self-sufficient quasars would no longer be sensitive to the Hubble tension. Any change of $\eta$ with redshift would indicate distance duality relation violation, assuming no redshift-dependent systematics (such as an evolving UV--X-ray relation \citep{Lusso_2025}) exist in either distance measurement.

Two further remarks can be made on this point. Firstly, identifying the sound horizon as the culprit for the Hubble tension at $3\sigma$ from other explanations (left panel of \autoref{fig:triangles}) requires the absolute scale of the SARM distances to be accurate to $\simeq|\alpha_0|/3\simeq2.6\%$. The systematic uncertainties of the BLR modelling, which are not included in our forecasts, would therefore be the angular diameter counterpart of the uncertainty in the SNIa anchor of the UV--X-ray distances.

Secondly, comparing the SARM scale with that of BAO (bottom edge of each triangle in \autoref{fig:triangles}) amounts, in essence, to a measurement of $H_0$ from SARM \citep{Wang:2019gaq, Li_2025}, which simulations suggest could reach $\simeq2\%$ with $\simeq60$ objects \citep{Songsheng_2021}. This is enough to identify the `odd one out' in terms of the triangle corners without any UV--X-ray distances; in other words, the Hubble tension could be addressed with SARM distances alone. What the self-sufficient quasar measurement adds is the comparison with the SNIa-based luminosity distances without assuming a cosmological model, which is what allows the distance duality relation itself to be tested.

\section{Conclusions}
\label{sec:conclusions}
Tests of the distance duality relation \citep{Etherington}, \autoref{eq:ddr}, usually combine luminosity and angular diameter distances from different probes, the most popular examples being SNIa and BAO. This requires matching the measurements, either by assuming a fiducial cosmology or by applying machine learning or Gaussian process methods, and inherently relies on the assumption of sky isotropy. To address these issues, we considered using quasars as self-sufficient probes, measuring both distances for the same object: the luminosity distance through the empirical UV--X-ray luminosity relation, and the angular diameter distance through spectroastrometry combined with reverberation mapping (SARM). Since both distances are measured for the same object, the constraints on the distance duality relation are independent of cosmological parameters, and the isotropy of any violation can be tested rather than assumed.

The two distances are at very different stages of maturity. Luminosity distances based on the UV--X-ray relation are an established cosmological probe, with thousands of observations \citep{Risaliti_2015,Lusso_2017,Risaliti_2019,Lusso_2020,Lusso_2025}, whereas SARM samples are more than two orders of magnitude smaller and confined to $z<0.2$ \citep{Wang:2019gaq,Li_2022,Li_2025}. Nevertheless, a single object is already enough to demonstrate the method: Mrk~509 ($z=0.0344$) \citep{Markaryan1973} is the only quasar that appears both in the UV--X-ray sample of \citet{Lusso_2020} and among the SARM quasars of \citet{Li_2025}, and combining the two measurements gives $\eta = 0.50^{+0.61}_{-0.27}$, consistent with the distance duality relation at $0.8\sigma$. Although dominated by uncertainties, this is, to our knowledge, the first test of the distance duality relation obtained from the luminosity and angular diameter distances of the same individual object.

Mrk~509 shows that the test can be performed, but not whether it can be competitive. To assess this, we forecast how many quasars would be required, and with what precision, to detect violations of the distance duality relation. As benchmarks we considered a constant offset (M1) and a linear evolution with redshift (M2), with the fiducial values found by \citet{Teixeira:2025czm} from BAO, SNIa, and CMB measurements, which are phenomenologically interesting due to their connection to the Hubble tension and to dynamical dark energy. We divided the forecasts into three scenarios according to the measurement uncertainties: \scenario1, \scenario2, and \scenario3. For M1, we found that 840, 414, and 274 quasars are needed, respectively, for a $3\sigma$ detection. Given that upgrades to the GRAVITY instrument are expected to provide hundreds of new measurements up to $z\sim2$--$3$ \citep{Amorim_2021, GRAVITYplusWhitePaper}, such a robust detection seems realistically feasible only in the \scenario3 scenario, with the caveat that we assumed the $d_{\rm A}$ uncertainties to be independent of redshift; modelling their redshift dependence is needed to refine the forecast, and we defer this to a future work.

Detecting a violation that evolves with redshift is considerably harder. For M2, with a hypothetical sample of $N=1000$ quasars, the case of no violation, $\alpha_0 = \alpha_1 = 0$, is excluded at the $3.4\sigma$, $5.2\sigma$, and $6.6\sigma$ levels in the \scenario1, \scenario2, and \scenario3 scenarios respectively, yet the strong anti-correlation of $\alpha_0$ and $\alpha_1$ means that neither parameter is individually inconsistent with zero. Detecting the redshift dependence itself would require $8\times10^4$ quasars in the \scenario1 scenario and $2.6\times10^4$ in the \scenario3 scenario. 

A constant offset of the size considered here is, by construction, the recalibration of the SNIa distance scale that restores concordance with BAO, and this is where quasars are particularly well placed. The absolute scale of BAO distances relies on the sound horizon, and therefore on early-time physics, whereas SARM distances are geometric and independent of it. A quasar measurement of $\eta$, with UV--X-ray distances anchored to the SNIa scale, can thus discriminate between a BAO-side origin of the Hubble tension, for which $\eta\simeq1$, and a SNIa-side origin, for which $\eta$ is consistent with the fiducial offset (\autoref{fig:triangles}). 

In the latter case, a SNIa miscalibration cannot be told apart from a genuine violation of the distance duality relation by a constant offset alone, and requires either an anchor independent of SNIa or the redshift dependence of $\eta$, which, unlike the constant offset, is insensitive to a constant calibration factor. The discrimination also requires the SARM distance scale to be accurate at the $\simeq2.6\%$ level; our forecasts for M1 assume the calibration uncertainties of both distance scales to be negligible. Although current SARM samples are far from the sizes we found necessary to detect a constant distance duality violation, the extension of SARM with GRAVITY+ could make quasars a complementary probe of the distance duality relation and, through it, of the calibration of the SNIa distance scale.

\acknowledgments
We thank Will Handley, Yan-Rong Li, Elisabeta Lusso and Sunny Vagnozzi for helpful discussions. D.P. acknowledges support from the Istituto Nazionale di Fisica Nucleare (INFN) through the Commissione Scientifica Nazionale 4 (CSN4) Iniziativa Specifica ``Quantum Fields in Gravity, Cosmology and Black Holes'' (FLAG). JLM acknowledges support from the ``la Caixa'' Foundation
(ID 100010434; fellowship code LCF\slash BQ\slash EU24\slash 12060091). NBH is supported by the research environment and infrastructure of the Handley Lab at the University of Cambridge. This publication is based upon work from the COST Action CA21136 ``Addressing observational tensions in cosmology with systematics and fundamental physics'' (CosmoVerse), supported by COST (European Cooperation in Science and Technology). We acknowledge the use of Large Language Models, specifically Claude Code Opus 5.5, and Codex Sol 5.5 and Astra 5.6, to adapt pre-existing forecasting code for use with quasars, and to help edit the manuscript. All code to reproduce the results and figures presented in this work can be found at \url{https://github.com/nataliehogg/quasar-ddr}.

\bibliography{ddr_forecast}

\end{document}